\documentclass[a4paper,11pt]{article}
\usepackage{jinstpub} 
\usepackage{lineno}
\usepackage{subfigure,dcolumn}
\usepackage{amsmath}
\usepackage{verbatim}
\usepackage{siunitx}
\usepackage{arydshln} 
\usepackage{threeparttable}
\usepackage{tabularx}
\usepackage{booktabs}
\usepackage{float}
\usepackage[colorlinks=true,linkcolor=blue,citecolor=blue,urlcolor=blue]{hyperref}

\title{Statistical research on determining sensitivity of neutrinoless double beta decays}

\author[1,a,b]{Haoyang Fu,\note{Co-corresponding author. (email:fu-hy21@mail.tsinghua.edu.cn)}}
\author[a,b,*]{Wentai Luo,}
\author[2,c]{Xiangpan Ji\note{Co-corresponding author. (email:jixp@mail.nankai.edu.cn)}}
\author[a,b,d]{and Shao-Min Chen}

\affiliation[a]{Department of Engineering Physics, Tsinghua University, Beijing, China}
\affiliation[b]{Center for High Energy Physics, Tsinghua University, Beijing, China}
\affiliation[c]{School of Physics, Nankai University, Tianjin, China}
\affiliation[d]{Key Laboratory of Particle $\&$ Radiation Imaging, Tsinghua University, Ministry of Education, Beijing, China}

\emailAdd{luowt@mail.tsinghua.edu.cn}

\abstract{
The determination of experimental sensitivity is a key step in the search for neutrinoless double beta decay ($0\nu\beta\beta$), providing a quantitative benchmark for detector design. Two commonly used statistical approaches are the counting method, which estimates sensitivity from the number of events in a predefined region of interest, and the fitting method, which extracts the signal contribution by fitting the full energy spectrum. In this work, we investigate both discovery sensitivity and exclusion sensitivity within these two approaches. Through statistical derivation and simulation verification, we show that the relative performance of the methods depends on both energy resolution and exposure, while at higher exposures the fitting method consistently yields more stringent sensitivity. These results provide guidance for selecting the optimal statistical method in future $0\nu\beta\beta$ experiments.}


\arxivnumber{2412.19859} 

\begin{document}
\maketitle
\flushbottom

\section{Introduction}
\label{sec:Introduction}
Neutrinoless double beta decay ($0\nu\beta\beta$), $(A,Z)\rightarrow (A,Z+2) + 2\mathrm{e}^-$, is a hypothetical decay process that could potentially occur in double beta decay isotopes. It is only allowed in the case of the Majorana neutrino but forbidden in the case of the Dirac neutrino~\cite{Dolinski:2019nrj}. Search for $0\nu\beta\beta$ plays a significant role in exploring physics beyond the Standard Model~\cite{Rodejohann:2011mu}.

Extensive experimental efforts are being dedicated to the search for $0\nu\beta\beta$~\cite{Zeng:2016pnx,CUORE:2022fgm,CUPID:2020aow,Chen:2022rzg,NnDEx-100:2023alw,GERDA:2020xhi,CDEX:2022bdk,PandaX:2023ggs,KamLAND-Zen:2022tow}. So far, no evidence has been observed. There are about 35 naturally occurring double beta decay isotopes, in which only those with a $Q$-value greater than \SI{2}{MeV} are adopted in the search. Depending on their physical and chemical properties, experimental designs generally fall into two main categories: crystal calorimeters and fluid calorimeters (including gaseous and liquid detectors). They differ significantly in energy resolution, background levels, and scalability. Consequently, different analysis approaches are adopted to set sensitivities on the $0\nu\beta\beta$ half-life. One is the counting method~\cite{CUORE:2022fgm,CUPID:2020aow,Chen:2022rzg,NnDEx-100:2023alw}, which involves statistical analysis of events within a predefined region of interest (RoI). The other one is the fitting method~\cite{GERDA:2020xhi,CDEX:2022bdk,PandaX:2023ggs,KamLAND-Zen:2022tow}, which utilizes spectrum fitting to calculate the sensitivity.

This research, with practical applications in the field, aims to compare the effectiveness of two methods quantitatively through statistical derivation and simulation verification. Section~\ref{sec:CountingAndFitting} outlines the fundamental principles, the connection between the two methods, and an analytical estimation. Section~\ref{sec:Simulation} presents a verification from the simulation. Section~\ref{sec:Discussion} discusses some issues of the fitting method, and Section~\ref{sec:Conclusion} concludes the paper.

\section{Principles of the counting and fitting methods}
\label{sec:CountingAndFitting}

In $0\nu\beta\beta$ experiments, the sensitivity to the half-life is typically of interest, and its calculation is equivalent to determining the sensitivity to the signal strength $s_{\mathrm{sens}}$. 

\begin{figure}
\centering
\subfigure[]{
\includegraphics[width=0.45\linewidth]{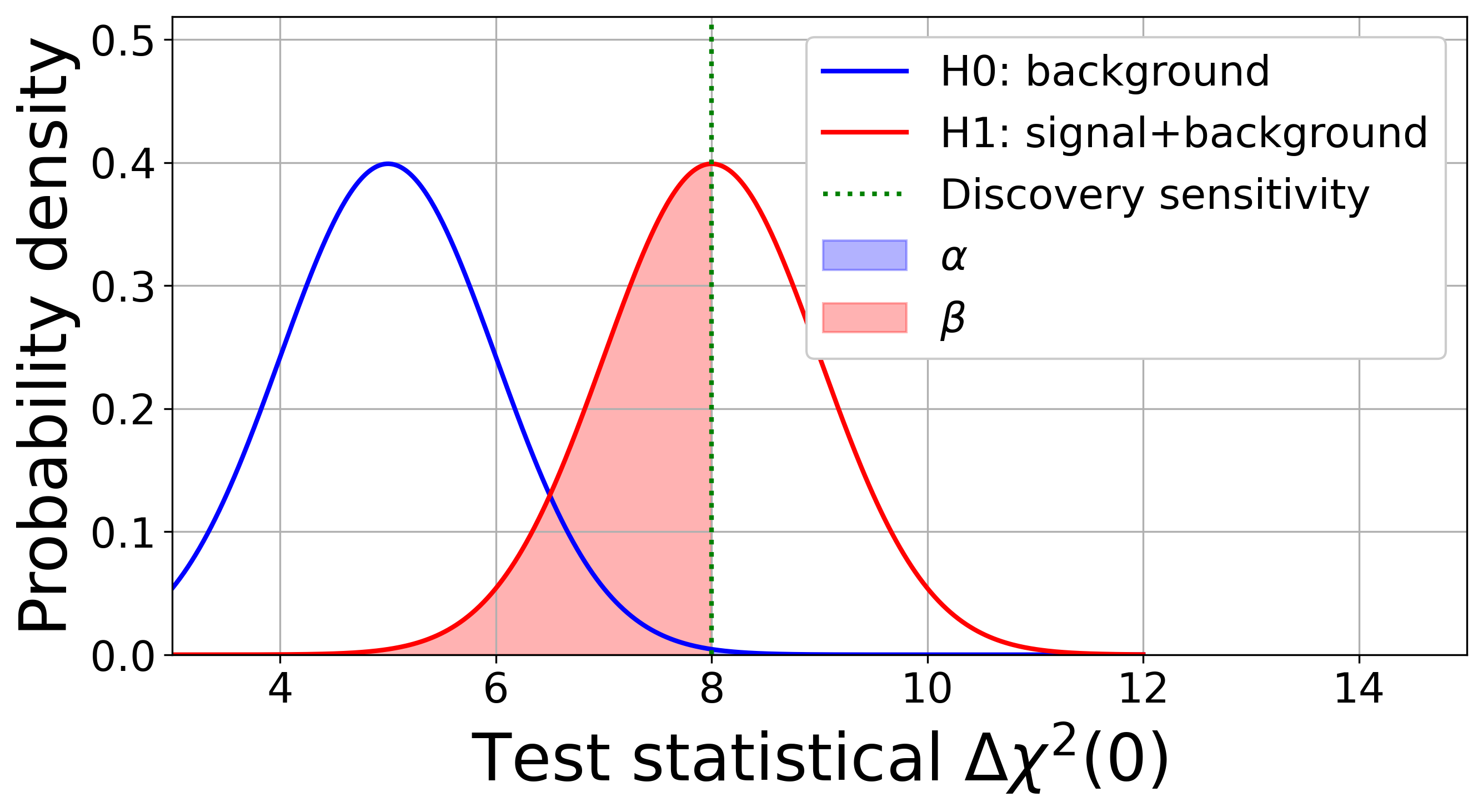}
}
\subfigure[]{
\includegraphics[width=0.45\linewidth]{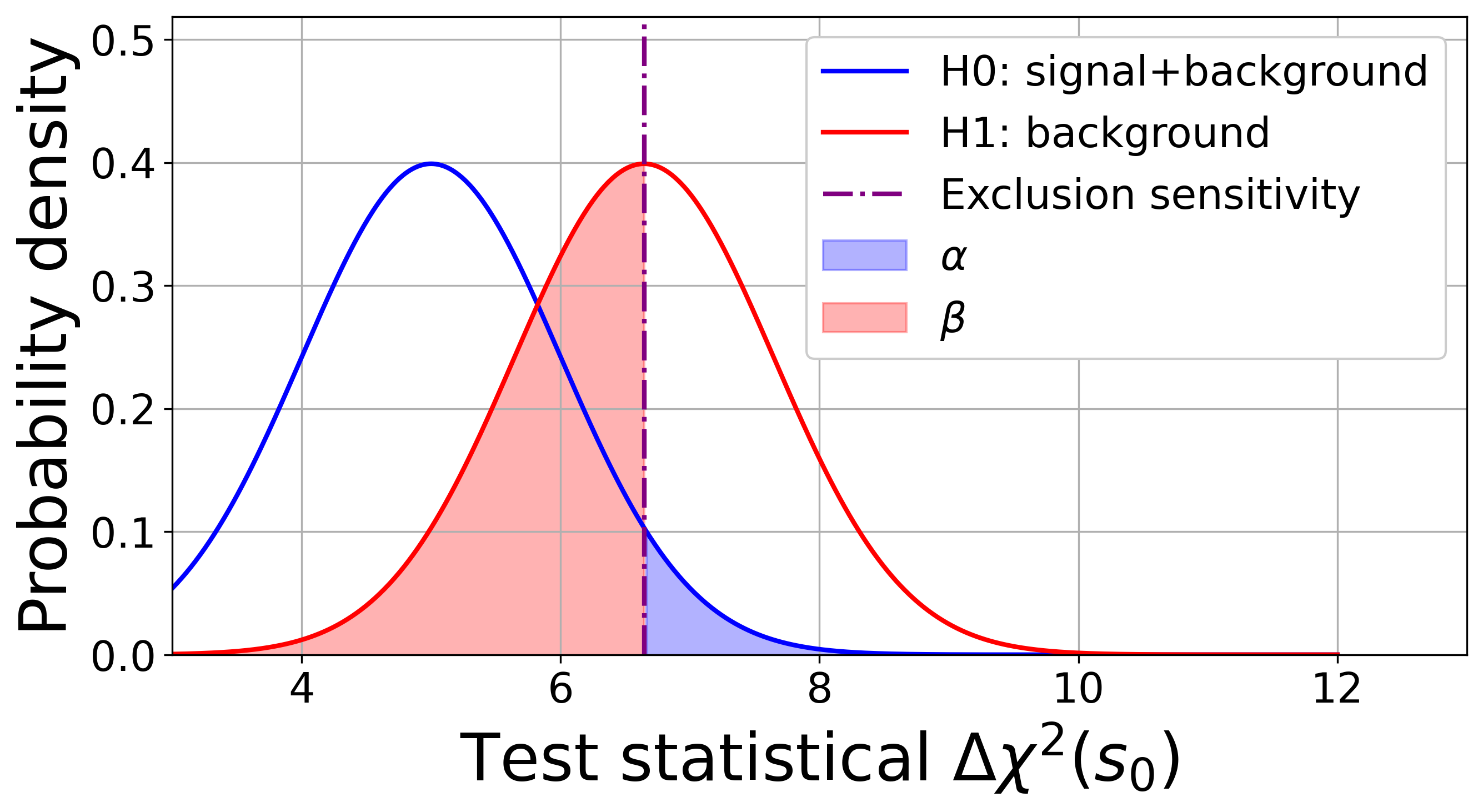}
}
\caption{Simple examples of calculating (a) discovery sensitivity and (b) exclusion sensitivity. The figures shows the test statistic distributions, which is defined differently in the two cases ($\Delta \chi^2(0)$ for discovery and $\Delta \chi^2(s_0)$ for exclusion); see Section~\ref{subsec:Counting} for details. The blue curve shows the $\Delta \chi^2$ probability densities under the null hypothesis $H_0$, which is background-only for discovery sensitivity and signal+background for exclusion sensitivity. The red curve shows the $\Delta \chi^2$ probability densities under the alternative hypothesis $H_1$, which is signal+background for discovery sensitivity and background-only for exclusion sensitivity. The blue shaded area denotes the type-I error $\alpha$. The red shaded area denotes the type-II error $\beta$, which is fixed at 0.5 in both cases.}\label{fig:H0H1}
\end{figure}

Generally, the term “sensitivity” is commonly used in two contexts: discovery sensitivity and exclusion sensitivity, with the assignment of null hypothesis $H_0$ and alternative hypothesis $H_1$ depending on context. For discovery sensitivity, one test $H_0:\, S=0$ (background only) against $H_1:\, S=s_0>0$ (signal+background). The discovery sensitivity $s_{\mathrm{disc}}$ is the signal strength required to reject $H_0$ at confidence level $1-\alpha$ with power $1-\beta=0.5$. For exclusion sensitivity, the roles are reversed. One test $H_0:\, S=s_0>0$ (signal+background) against $H_1:\, S=0$ (background only), and the exclusion sensitivity is the value of $s_0$ that would be excluded at confidence level $1-\alpha$ in 50$\%$ of background-only experiments (i.e., $1-\beta=0.5$). In this study, we examine both definitions and compare the counting and fitting methods using these two metrics. Fig.~\ref{fig:H0H1} shows a simple example.

\subsection{Counting method}
\label{subsec:Counting}

The counting method needs to define a region of interest (RoI) for the calculation of $s_{\mathrm{sens}}$. Assuming $N_{\mathrm{pred}}$ and $N_{\mathrm{meas}}$ as the predicted and measured number of events within the RoI, respectively, we can write Pearson’s chi-square and Neyman's chi-square as follows~\cite{Ji:2019yca}:

\begin{align}
\chi^2_{\mathrm{Pearson}} &= 
\frac{(N_{\mathrm{pred}} - N_{\mathrm{meas}})^2}{N_{\mathrm{pred}}}, \\
\chi^2_{\mathrm{Neyman}} &= 
\frac{(N_{\mathrm{pred}} - N_{\mathrm{meas}})^2}{N_{\mathrm{meas}}}.
\end{align}
We also denote $B_{\mathrm{pred}}$ and $S$ as the predicted total background and signal counts in the RoI, giving $N_\mathrm{pred}=S + B_{\mathrm{pred}}$.

\subsubsection{Discovery sensitivity for counting method}
\label{subsubsec:counting disc}
We define the test statistic $\Delta \chi^2(S=0)$ as
\begin{equation}
    \label{eq:deltaChi2_counting}
    \Delta \chi^2(S=0) = \chi^2_{\mathrm{Pearson}}(S=0) -\chi^2_{\mathrm{Pearson}}(\hat{S}),
\end{equation}
where $\hat{S}$ denotes the $S$ that minimizes the Pearson's chi-square value. 

Under the null hypothesis $H_0:S=0$, we denote $\Delta\chi^2({S=0})$ as $\Delta\chi^2(S=0\mid S=0)$. Here, the first “$S=0$” specifies the parameter value in the test statistic $\Delta\chi^2(S)$, whereas the second “$S=0$” denotes that the pseudo-data are taken under the discovery sensitivity null hypothesis $H_0:S=0$. In this case, $\Delta\chi^2(S=0\mid S=0)$ follows that~\cite{Cowan:2010js}:
\begin{equation}
    \Delta\chi^2(S=0\mid S=0) \sim \frac{1}{2} \delta(0) +  \frac{1}{2} \chi^2(1),
\end{equation}
where the $\delta(0)$ represents a Dirac delta function at zero. The term $\frac{1}{2} \delta(0)$ arises due to the physical boundary condition imposed in the fit, namely that the signal strength must satisfy $s \ge 0$. As a result, for those samples whose fits lie at the boundary, the test statistic takes the value $\Delta\chi^2(S=0\mid S=0) = \chi^2_{\mathrm{Pearson}}(S=0) -\chi^2_{\mathrm{Pearson}}(\hat{S}=0)=0$.

The cumulative distribution function of $\Delta\chi^2(S=0\mid S=0)$ is given by~\cite{Cowan:2010js}:
\begin{equation}
    F(\Delta\chi^2(S=0\mid S=0))=\Phi\left(\sqrt{\Delta\chi^2(S=0\mid S=0)}\right),
\end{equation}
where $\Phi$ is cumulative Gaussian distribution. Therefore, the $\alpha$-quantile of this mixed $\chi^2$ distribution, namely $\Delta \chi^2_{\alpha}(S=0\mid S=0)$, corresponds to $\Delta \chi^2_{\alpha}(S=0\mid S=0)=z_\alpha^2$, where $z_\alpha$ is the one-sided $\alpha$-quantile of the standard normal distribution (i.e. $z_\alpha = \Phi^{-1}(1-\alpha)$).

Under the alternative hypothesis $H_1: S = s_0$, the cumulative distribution function of $\Delta \chi^2(S=0)$ (denoted as $\Delta \chi^2 (S=0\mid s_0)$) is given by~\cite{Cowan:2010js}:
\begin{equation}
\label{eq:S0H1}
    F(\Delta \chi^2 (S=0\mid s_0))=\Phi\left(\sqrt{\Delta \chi^2 (S=0\mid s_0)}-\frac{\hat{S}}{\sigma_S}\right),
\end{equation}
where $\sigma_S$ is the standard deviation of $\hat{S}$, and it can be evaluated by the Asimov value $\Delta \chi^2_{\mathrm{Asimov}}(S=0\mid s_0)$ as~\cite{Cowan:2010js}:

\begin{equation}
\label{eq:SimgaS_Count_Disc}
    \sigma_S^2 =\frac{s_0^2}{\Delta \chi^2_{\mathrm{Asimov}}(S=0\mid s_0)}.
\end{equation}

When the Asimov dataset is employed, we have $N_{\mathrm{meas}} = B_{\mathrm{pred}} + S =B_{\mathrm{pred}} + s_0$, which leads to:
\begin{equation}
    \Delta \chi^2_{\mathrm{Asimov}}(S=0\mid s_0) = \frac{s_0^2}{B_{\mathrm{pred}}},
\end{equation}
which gives:
\begin{equation}
    \sigma_S=\sqrt{B_\mathrm{pred}}.
\end{equation}

Therefore, combining with Equation~\ref{eq:S0H1}, we obtain that the median of $\Delta \chi^2 (S=0\mid s_0)$ is given by
\begin{equation}
    \Delta \chi^2_{\mathrm{median}}(S=0\mid s_0)=\frac{s_0^2}{B_{\mathrm{pred}}}.
\end{equation}

The discovery sensitivity $s_{\mathrm{disc}}$ is therefore defined as the value of $s_0$ that satisfies the condition $\Delta \chi^2_{\mathrm{median}}(S=0\mid s_0) = \Delta\chi^2_\alpha(S=0\mid S=0) = z_\alpha^2$, which leads to:

\begin{equation}
    \label{eq:sens}
    s_{\mathrm{disc}}=z_\alpha \sqrt{B_\mathrm{pred}}.
\end{equation}

$s_{\mathrm{disc}}$ corresponds to a lower limit on the total number of signal events, which is related to the half-life sensitivity by:
\begin{equation}
    \label{eq:trans_disc}
    T_{1/2}^{\mathrm{disc}} = \frac{N_A \times \ln2}{s_{\mathrm{disc}}} \frac{x\eta}{A_{\mathrm{iso}}} \times Mt\times \epsilon,
\end{equation}
where $N_A$ is Avogadro's number, $x$ is the loading mass fraction of the $0\nu\beta\beta$ isotopes and $\eta$ is the isotopic abundance, $A_{\mathrm{iso}}$ is the atomic number of the isotope, $M$ is the detector mass, $t$ is the detector run time, and $\epsilon$ represents the signal selection efficiency within the RoI.

Assuming $B$ is the background number per unit mass, time and energy, and $\Delta$ is the width of the RoI, we can write $B_{\mathrm{pred}}$ as:

\begin{equation}
    B_{\mathrm{pred}} =B \cdot \Delta Mt,
\end{equation}
which allows us to rewrite the half-life sensitivity in a more commonly used form~\cite{DellOro:2016tmg}, which is often used to estimate the sensitivity:
\begin{equation}
\label{eq:Tsensitivity}
T_{1/2}^{\mathrm{disc}} = \frac{N_A \times \ln2}{z_\alpha} \frac{x\eta}{A_{\mathrm{iso}}} \times \sqrt{\frac{M\cdot t}{B \cdot \Delta}} \times \epsilon.
\end{equation}

\subsubsection{Exclusion sensitivity for counting method}
\label{subsubsec:Counting median}

For the calculation of exclusion sensitivity, we aim to keep the final expression in a form analogous to Eq.~\ref{eq:Tsensitivity}. Therefore, we define the exclusion sensitivity test statistic $\Delta\chi^2(S=s_0)$ as:
\begin{align}
\label{eq:deltaChi2_counting_med}
\Delta\chi^2(S=s_0) &= 
\chi^2_{\mathrm{Neyman}}\big(S=s_0) 
 - \chi^2_{\mathrm{Neyman}}\big(\hat{S}\big).
\end{align}

Similarly, under the null hypothesis for exclusion sensitivity $H_0: S = s_0$, the cumulative distribution function of $\Delta\chi^2(S=s_0)$ (denoted as $\Delta \chi^2(S=s_0\mid s_0)$) is given by~\cite{Cowan:2010js}:
\begin{equation}
F(Q) =
\begin{cases}
2\Phi(\sqrt{Q}) - 1, & Q \le \tfrac{s_0^2}{\sigma_S^2}, \\ [6pt]
\Phi(\sqrt{Q}) + \Phi!\left(\tfrac{Q + s_0^2/\sigma_S^2}{2s_0/\sigma_S}\right) - 1,
& Q > \tfrac{s_0^2}{\sigma_S^2},
\end{cases}
\end{equation}
where we define
\begin{equation}
Q \equiv \Delta \chi^2(S=s_0\mid s_0).
\end{equation}

Analogous to Equation~\ref{eq:SimgaS_Count_Disc}, we can compute $\sigma_S$ in this case as
\begin{equation}
    \sigma_S = \frac{s_0}{\sqrt{\Delta \chi^2_{\mathrm{Asimov}}(S=s_0\mid s_0)}}.
\end{equation}
Moreover, we obtain
\begin{equation}
    \Delta \chi^2_{\mathrm{Asimov}}(S=s_0\mid s_0)=0.
\end{equation}

This implies that in this case one always has
\begin{equation}
    F(\Delta \chi^2(S=s_0\mid s_0)) = 2\Phi\!\left(\sqrt{\Delta \chi^2(S=s_0\mid s_0)}\right) - 1,
\end{equation}
and the $\alpha$-quantile of $\Delta\chi^2(S=s_0\mid s_0)$ should be:
\begin{equation}
    \Delta \chi^2_{\mathrm{\alpha}}(S=s_0\mid s_0)=z_{\alpha/2}^2.
\end{equation}

Under the alternative hypothesis $H_1: S = 0$, we obtain~\cite{Cowan:2010js}:
\begin{align}
\label{eq:S1H0}
F(Q') &= \Phi\!\Big( \sqrt{Q'} + \frac{s_0}{\sigma_S}\Big) + 
\begin{cases} 
\Phi\!\Big( \sqrt{Q'} - \frac{s_0}{\sigma_S}\Big)-1, & 
Q' \le s_0^2 / \sigma_S^2, \\[2mm]
\Phi\!\Big( \frac{Q' - s_0^2/\sigma_S^2}{2s_0/\sigma_S}\Big)-1, & 
Q' > s_0^2 / \sigma_S^2,
\end{cases}
\end{align}
where we define
\begin{equation}
Q' \equiv \Delta \chi^2(S=s_0\mid S=0).
\end{equation}

The rejection region of the signal, $\mathcal{R}$, is defined as
\begin{equation}
    \mathcal{R}=\{s_0\mid  \Delta \chi^2(S=s_0\mid S=0) \ge\Delta \chi^2_{\mathrm{\alpha}}(S=s_0\mid s_0)\}.
\end{equation}

Using the Asimov dataset, we have
\begin{equation}
    \Delta \chi^2_{\mathrm{Asimov}}(S=s_0\mid S=0)=\frac{s_0^2}{B_\mathrm{pred}},
\end{equation}
which leads to
\begin{equation}
    \sigma_S = \sqrt{B_\mathrm{pred}}.
\end{equation}

To obtain the sensitivity $s_\mathrm{excl}$, one needs to substitute $\sigma_S$ into Equation~\ref{eq:S1H0}, compute the median of the corresponding distribution, and then determine the value of $s_0$ such that
\begin{equation}
    \Delta \chi^2_{\mathrm{median}}(S=s_0\mid S=0) =\Delta \chi^2_{\mathrm{\alpha}}(S=s_0\mid s_0).
\end{equation}
In general, this can only be solved numerically.

It should be noted that the boundary of the rejection region is located at $\Delta \chi^2(S=s_0\mid S=0) = \Delta \chi^2_{\mathrm{\alpha}}(S=s_0\mid s_0) = z^2_{\alpha/2}$. For most practical cases, one has
\begin{equation}
    \Phi\!\left( \sqrt{\Delta \chi^2(S=s_0\mid S=0)} + \tfrac{s_0}{\sigma_S}\right)\approx 1.
\end{equation}
For example, when $\alpha=0.05$, we have 
\begin{equation}
    \Phi\!\left( \sqrt{\Delta \chi^2(S=s_0\mid S=0)} + \tfrac{s_0}{\sigma_S}\right) > 0.975.
\end{equation}

Therefore, as an approximation, we may take
\begin{equation}
    F(\Delta \chi^2(S=s_0\mid S=0)) \approx \Phi\!\left( \sqrt{\Delta \chi^2(S=s_0\mid S=0)} - \frac{s_0}{\sigma_S}\right),
\end{equation}
or
\begin{equation}
    F(\Delta \chi^2(S=s_0\mid S=0)) \approx \Phi\!\left( \frac{\Delta \chi^2(S=s_0\mid S=0) - s_0^2/\sigma_S^2}{2s_0/\sigma_S}\right),
\end{equation}
both of which lead to the same median:
\begin{equation}
    \Delta \chi^2_{\mathrm{median}}(S=s_0\mid S=0)= \frac{s_0^2}{\sigma_S^2}.
\end{equation}

Consequently, the exclusion sensitivity is given by
\begin{equation}
    s_{\mathrm{excl}}=z_{\alpha/2} \sqrt{B_\mathrm{pred}}.
\end{equation}
Similarly, this can be translated into the exclusion sensitivity of half-life:
\begin{equation}
    T_{1/2}^{\mathrm{excl}} = \frac{N_A \times \ln2}{z_{\alpha/2}} \frac{x\eta}{A_{\mathrm{iso}}} \times \sqrt{\frac{M\cdot t}{B \cdot \Delta}} \times \epsilon.
\end{equation}

\subsection{Fitting method}
\label{subsec:Fitting}
The fitting method typically employs a binned likelihood function, using a profiled-likelihood ratio test to test the null hypothesis. A typical bin-wise maximum likelihood function is the product of Poisson probabilities for all bins:
\begin{equation}
\mathcal{L}(\hat{S},\{\hat{B}_{\mathrm{pred}}\}) = \prod_{i} \frac{\nu_i^{n_i} e^{-\nu_i}}{n_i!},
\end{equation}
where $n_i$ and $\nu_i$ are the measured and predicted number of events in the $i$-th bin. The predicted experimental background intensity $\{B_{\mathrm{pred}}\}$, as well as the predicted signal intensity $S$, are implicit in $\nu_i$: we assume the background and signal spectra are well known as $f_j(x)$ and $f_s(x)$, where $j$ denotes the index of background. Then, we have $\nu_i=\int_{i-\mathrm{th\ bin}}^{}{\left(\sum_j f_j(x) B^j_{\mathrm{pred}} + f_s(x)S \ \right) dx}$. $\{\hat{B}_{\mathrm{pred}}\}$ and $\hat{S}$ represent the values of $\{B_{\mathrm{pred}}\}$ and $S$ that maximize $\mathcal{L}$, respectively.

\subsubsection{Discovery sensitivity for fitting method}
\label{subsubsec:fitting disc}

The likelihood ratio $\Lambda_{\mathrm{disc}}$ for determining the discovery sensitivity is defined as:
\begin{equation}
    \Lambda(S=0) = \frac{\mathcal{L}(S=0,\{\hat{\hat{B}}_{\mathrm{pred}}\})}{\mathcal{L}(\hat{S},\{\hat{B}_{\mathrm{pred}}\})},
\end{equation}
where $\{\hat{\hat{B}}_{\mathrm{pred}}\}$ represents the value of $\{B_{\mathrm{pred}}\}$ that maximizes $\mathcal{L}$ when $S=0$ is fixed.

For $H_0:S=0$, the likelihood ratio described above satisfies~\cite{Cowan:2010js}:
\begin{equation}
    \Delta \chi^2(S=0 \mid S=0) = -2\ln{\Lambda
    (S=0)} \sim \frac{1}{2}\delta(0) +  \frac{1}{2}\chi^2(1),
\end{equation}
and for $H_1:S=s_0$, the $\Delta \chi^2(S=0)$ is denoted as $\Delta \chi^2_{S=0|H_1}$.

It is also common to add constraints on the background strength into the test statistic in the form with penalty terms, such that $\Delta\chi^2 = -2\ln{\Lambda} + \chi^2_{\mathrm{penal}}$, where:
\begin{equation}
\chi^2_{\mathrm{penal}}=\sum_j\left(\frac{B_j^{\mathrm{pred}}-B^{\mathrm{meas}}_j}{\sigma_j}\right)^2,
\end{equation}
$B_j^{\mathrm{pred}}$ denotes the predicted strength of the $j$-th background component, which is the fit parameter, while $B^{\mathrm{meas}}_j$ and $\sigma_j$ represent the measured value and the standard deviation of the $j$-th background strength, respectively. The inclusion of these penalty terms does not affect the subsequent derivations or conclusions~\cite{Cowan:2010js}.

\begin{figure}
\centering
\includegraphics[width=0.7\linewidth]{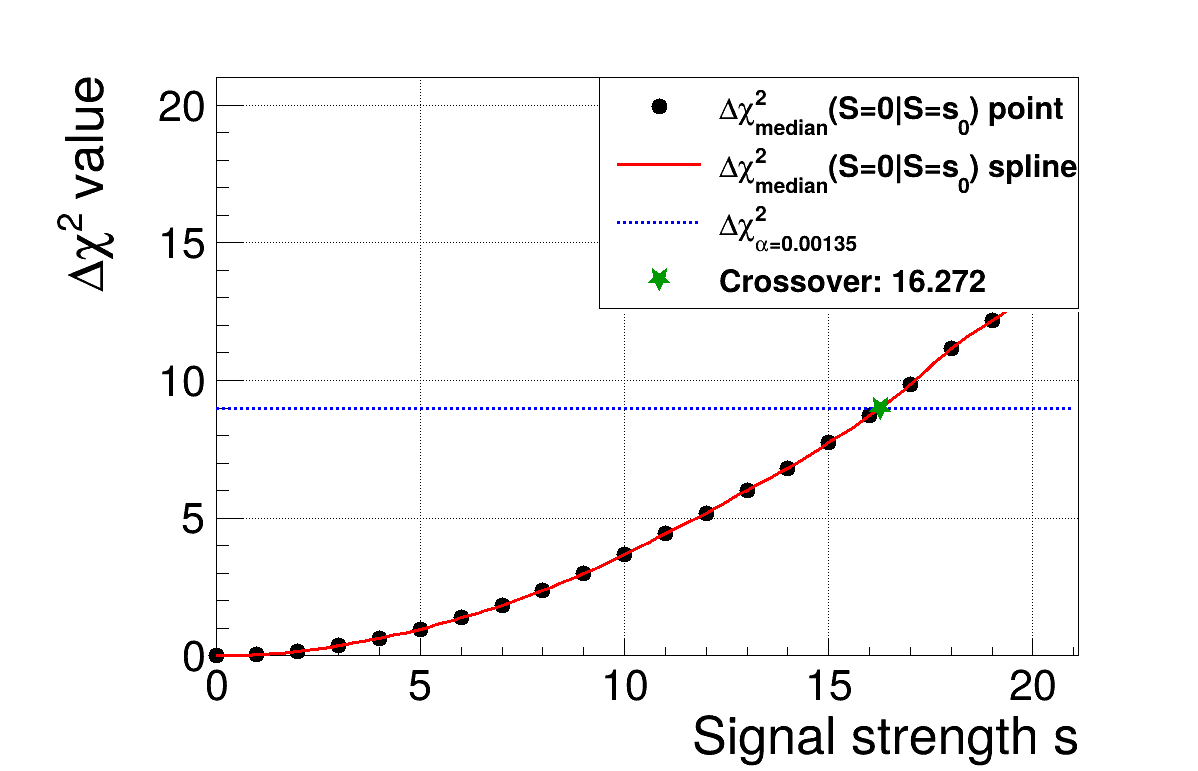}
\caption{The 3$\sigma$ discovery sensitivity analysis process of the fitting method. The figure is obtained from a liquid scintillator detector doped $^{130}$Te with an energy resolution of 3$\%$. The blue dashed line corresponds to $\Delta \chi^2_{\alpha}(S=0\mid S=0)$, and the black dots are $\Delta \chi^2_{\mathrm{median}}(S=0\mid s_0)$ at different $s_0$ values. The red curve shows a cubic spline interpolation of $\Delta \chi^2_{\mathrm{median}}(S=0\mid s_0)$. The green star represents the crossover point of $\Delta \chi^2_{\alpha}(S=0\mid S=0)$ and $\Delta \chi^2_{\mathrm{median}}(S=0\mid s_0)$, which indicates the discovery sensitivity $s_{\mathrm{disc}}$ result.}\label{fig:q0qs_example}
\end{figure}

With the $\Delta \chi^2(S=0)$ as the test statistic, similar to Section~\ref{subsubsec:counting disc}, $s_{\mathrm{disc}}$ is defined as the value of $s_0$ that makes $\Delta \chi^2_{\alpha}(S=0\mid S=0) = \Delta \chi^2_{\mathrm{median}}(S=0\mid s_0)$ at a given confidence level $1-\alpha$. A numerical evaluation can be performed in exactly the same manner as in Section~\ref{subsubsec:counting disc}. However, in practice, it is more common to employ toy Monte Carlo (toy MC) simulations to obtain the $\alpha$-quantile of $\Delta \chi^2(S=0\mid S=0)$ and the median of $\Delta \chi^2 (S=0\mid s_0)$. Fig.~\ref{fig:q0qs_example} shows an example of this toy MC process, obtained from a liquid scintillator detector doped $^{130}$Te which will be illustrated in Section~\ref{sec:Simulation}, with an energy resolution of 3$\%$ (defined by the standard deviation $\sigma_{\mathrm{E}}$ at \SI{1}{MeV} as $\frac{\sigma_{\mathrm{E}}}{1\ \mathrm{MeV}}$).

\subsubsection{Exclusion sensitivity for fitting method}
For the exclusion sensitivity, we define the likelihood ratio as
\begin{equation}
    \Lambda(S=s_0) = \frac{\mathcal{L}(S=s_0,\{\hat{\hat{B}}_{\mathrm{pred}}\})}{\mathcal{L}(\hat{S},\{\hat{B}_{\mathrm{pred}}\})},
\end{equation}
and the corresponding test statistic $\Delta\chi^2(S=s_0)$ as
\begin{equation}
    \Delta\chi^2(S=s_0)=-2\ln \Lambda(S=s_0).
\end{equation}

\begin{figure}
\centering
\includegraphics[width=0.7\linewidth]{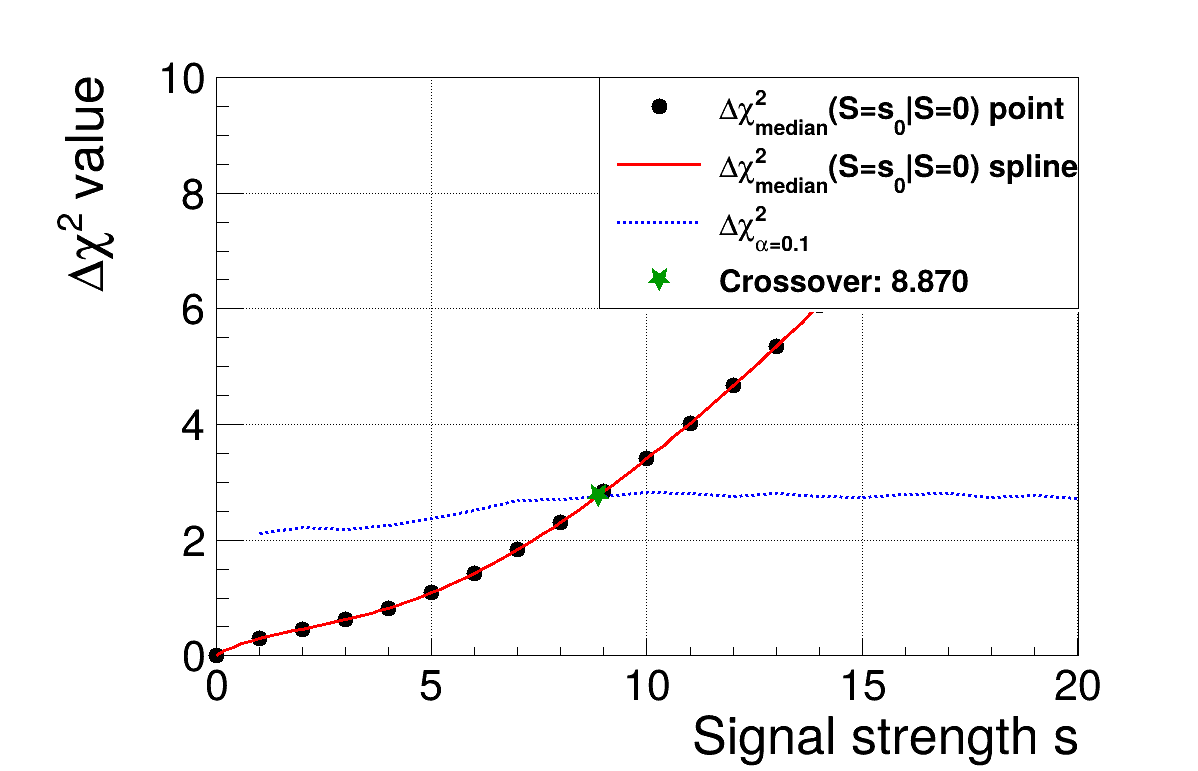}
\caption{The 90$\%$ C.L. exclusion sensitivity analysis process of the fitting method. The figure is obtained from a liquid scintillator detector doped $^{130}$Te with an energy resolution of 3$\%$. The blue dashed curve corresponds to $\Delta \chi^2_{\alpha}(S=s_0\mid s_0)$, and the black dots are $\Delta \chi^2_{\mathrm{median}}(S=s_0\mid S=0)$ at different $s_0$ values. The red curve shows a cubic spline interpolation of $\Delta \chi^2_{\mathrm{median}}(S=s_0\mid S=0)$. The green star represents the crossover point of $\Delta \chi^2_{\alpha}(S=s_0\mid s_0)$ and $\Delta \chi^2_{\mathrm{median}}(S=s_0\mid S=0)$, which indicates the exclusion sensitivity $s_{\mathrm{excl}}$ result.}\label{fig:q0qs_example_med}
\end{figure}

Similar to Section~\ref{subsubsec:Counting median}, the value of $s_0$ satisfying
\begin{equation}
\Delta \chi^2_{\alpha}(S=s_0\mid s_0) = \Delta \chi^2_{\mathrm{median}}(S=s_0\mid S=0)    
\end{equation}
can be determined numerically in the same manner, or equivalently, via toy Monte Carlo (toy MC) simulations. Fig.~\ref{fig:q0qs_example_med} shows an example of this toy MC procedure, obtained for a liquid scintillator detector doped with $^{130}$Te which will be illustrated in Section~\ref{sec:Simulation}, with an energy resolution of 3$\%$.

\subsection{Connection between the two methods}
\label{subsec:Correlation}
To see clearly the connection between the two methods, we can process the $\chi^2_{\mathrm{poisson}}=-2\mathcal{L}$ in fitting method as follows, dividing it into two parts:

\begin{align}
    \chi^2_\mathrm{Poisson} & = 2 \sum_{i} \left( \nu_{i} - n_i + n_i \log \frac{n_i}{\nu_i} \right)\notag \\
    & = 2 \left( N_{\text{pred}} - N_{\text{meas}} + N_{\text{meas}} \log \frac{N_{\text{meas}}}{N_{\text{pred}}} \right) + 2 \sum_{i} n_i \log \left(\frac{n_i}{\nu_i}\right) \cdot \frac{\nu_i}{n_i},
\end{align}
where $n_i$ and $v_i$ represent the measured and predicted number of events within the $i$-th bin, $N_{\mathrm{meas}} = \sum_i n_i$ and $N_{\mathrm{pred}} = \sum_i v_i$ represent the total number of measured and predicted events within the fitting range, respectively. After simplification, we obtain:
\begin{align}
    \chi^2_\mathrm{Poisson} ={} & 2\left(N_{\mathrm{meas}} 
        \ln \frac{N_{\mathrm{meas}}}{N_{\mathrm{pred}}}
        + N_{\mathrm{pred}} - N_{\mathrm{meas}}\right) + 2\sum_i \left(
        n_i \ln \frac{n_i}{
        v_i \cdot \frac{N_{\mathrm{meas}}}{N_{\mathrm{pred}}}}
        \right),
\end{align}
The right-hand side of the equation is divided into two terms: the first one represents the event counting term, and the second one represents the energy spectrum term.

If we construct a test statistic that includes only the event counting term, the fit requires only a single parameter $N_{\mathrm{pred}} = \sum_j B^j_{\mathrm{pred}} + S$. Then the likelihood function $\mathcal{L}$ corresponds to the probability density function of the Poisson distribution as $\mathcal{L}(N_{\mathrm{pred}})=\frac{(N_{\mathrm{pred}})^{N_{\mathrm{meas}}} \cdot e^{-N_{\mathrm{pred}}}}{N_{\mathrm{meas}}!}$. Accordingly, we can define the Poisson-likelihood chi-square as:

\begin{align}
\chi^2_{\mathrm{Poisson}} 
    &= -2\ln \mathcal{L} = 2\left( N_{\mathrm{meas}} \ln \frac{N_{\mathrm{meas}}}{N_{\mathrm{pred}}}
        + N_{\mathrm{pred}} - N_{\mathrm{meas}} \right).
\end{align}

In analogy to Equation~\ref{eq:deltaChi2_counting} and Equation~\ref{eq:deltaChi2_counting_med}, a test statistic $\Delta \chi^2$ based on the likelihood ratio can be defined as:

\begin{align}
    \Delta \chi^2(S=0) 
        &= \chi^2_{\mathrm{Poisson}}(N_{\mathrm{pred}}=\sum_j B_j^{\mathrm{pred}}) 
         - \chi^2_{\mathrm{Poisson}}(\hat{N}_{\mathrm{pred}}) \nonumber \\
        &= \chi^2_{\mathrm{Poisson}}(N_{\mathrm{pred}}=\sum_j B_j^{\mathrm{pred}}),
\end{align}
and
\begin{align}
    \Delta\chi^2(S=s_0) 
        &= \chi^2_{\mathrm{Poisson}}(N_{\mathrm{pred}}=\sum_j B_j^{\mathrm{pred}}+s_0) 
         - \chi^2_{\mathrm{Poisson}}(\hat{N}_{\mathrm{pred}}) \nonumber \\
        &= \chi^2_{\mathrm{Poisson}}(N_{\mathrm{pred}}=\sum_j B_j^{\mathrm{pred}}+s_0),
\end{align}
where $\hat{N}_{\mathrm{pred}}$ denotes the value of $N_{\mathrm{pred}}$ that maximizes $\mathcal{L}$.

As discussed in Ref.~\cite{Ji:2019yca}, the quantity $\chi^2_{\mathrm{Poisson}}$ above corresponds to the Poisson-likelihood chi-square function, whereas $\chi^2_{\mathrm{Pearson}}$ in Equation~\ref{eq:deltaChi2_counting} is Pearson’s chi-square function and $\chi^2_{\mathrm{Neyman}}$ in Equation~\ref{eq:deltaChi2_counting_med} is Neyman’s chi-square function. These three forms yield comparable results in the large-statistics limit.

The connection implies that the performance of the counting method should be close to that of the fitting method in which the event counting term is used to construct the chi-square function.

\section{Simulation-based comparison}
\label{sec:Simulation}
As an application example and a verification, we conduct a simulation study and compare the counting and fitting methods for the sensitivity determination of $0\nu\beta\beta$ half-life for a liquid scintillator detector doped with 1$\%$ natural Te. 

\subsection{Detector configuration and background modeling}
\label{subsec:Backgrounds}
The simulation assumes a 500-ton liquid scintillator detector with 1$\%$ by mass of natural Te isotopes loaded, which corresponds to 1.71 tons of $^{130}$Te.

In the simulation, we consider two irreducible background sources: two-neutrino double beta decay ($2\nu\beta\beta$) and the solar neutrino elastic scattering. The $2\nu\beta\beta$ background is intrinsic and modeled using an analytical spectral shape~\cite{Zuber:2020kci}. More refined calculations of the $2\nu\beta\beta$ spectrum indicate only minor shape modifications, which do not affect the conclusions of this work~\cite{Simkovic:2018rdz}. For solar neutrinos, only $^8$B neutrinos can contribute due to their higher energies ~\cite{Serenelli:2016dgz}. The calculated recoil electron spectrum from elastic scattering take into account both theoretical fluxes and cross-sections~\cite{Bilenky:2005mx,Baudis:2013qla}. 

All background spectra are convolved with a Gaussian energy resolution and used to generate toy MC samples for sensitivity studies. 

\subsection{Comparison between the two methods}
\label{subsec:SimulationResult}
The counting and fitting methods are compared by the obtained 3$\sigma$ discovery sensitivities and 90$\%$ C.L. exclusion sensitivities of the $0\nu\beta\beta$ half-life. The comparison was performed for the cases with different energy resolutions. 

\begin{figure}
\centering
\subfigure[]{
\includegraphics[width=0.45\linewidth]{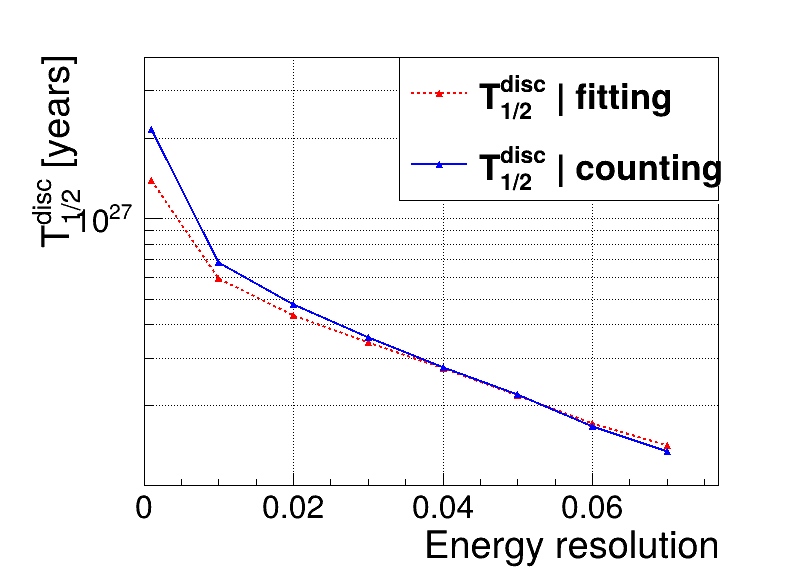}
}
\subfigure[]{
\includegraphics[width=0.45\linewidth]{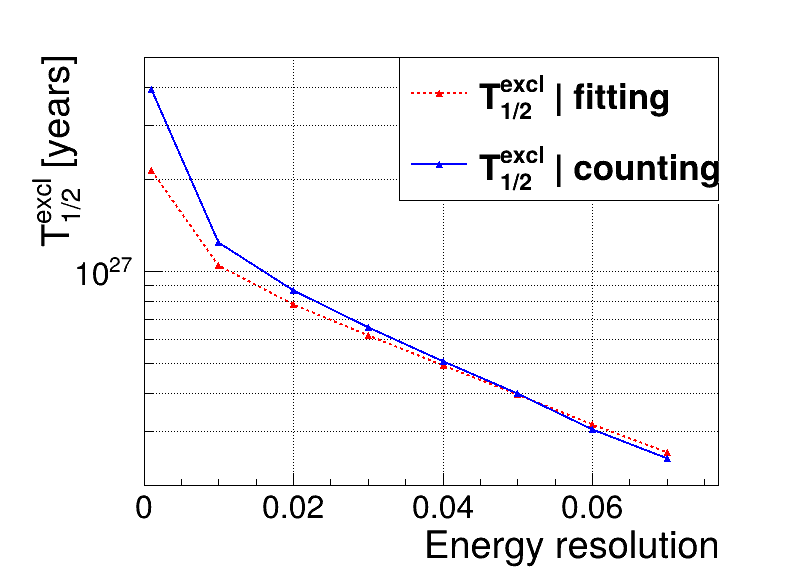}
}
\caption{The (a) 3$\sigma$ discovery sensitivities and (b) 90$\%$ C.L. exclusion sensitivities of $0\nu\beta\beta$ half-life obtained with $^{130}$Te by two methods. The detector run time is set to one year, corresponding to a $^{130}$Te exposure of 1.71 ton-year. The energy resolution ranges from 0.1$\%$ to 7$\%$.}\label{fig:compare_iso}
\end{figure}

In the simulation, the fitting procedure includes a penalty term for each background component, where $B^{\mathrm{meas}}_j$ is set to the expected value of its strength, and the corresponding uncertainty is taken as $\sigma_j=\sqrt{B^{\mathrm{meas}}_j}$. The half-life sensitivities for the counting method are derived using optimal RoIs, which vary with the energy resolution. As a result, Figure~\ref{fig:compare_iso} compares the $0\nu\beta\beta$ half-life discovery and exclusion sensitivities at a fixed one-year live time (corresponding to a $^{130}$Te exposure of 1.71 ton-year). With improved energy resolution, the results obtained from the counting method surpass those from the fitting method. This trend changes with increasing exposure: once the exposure becomes sufficiently large, the fitting method consistently outperforms the counting method. We will discuss this point in detail in Section~\ref{subsec:Fitting_exposure}.

\begin{figure}[]
  \centering
  \includegraphics[width=0.7\linewidth]{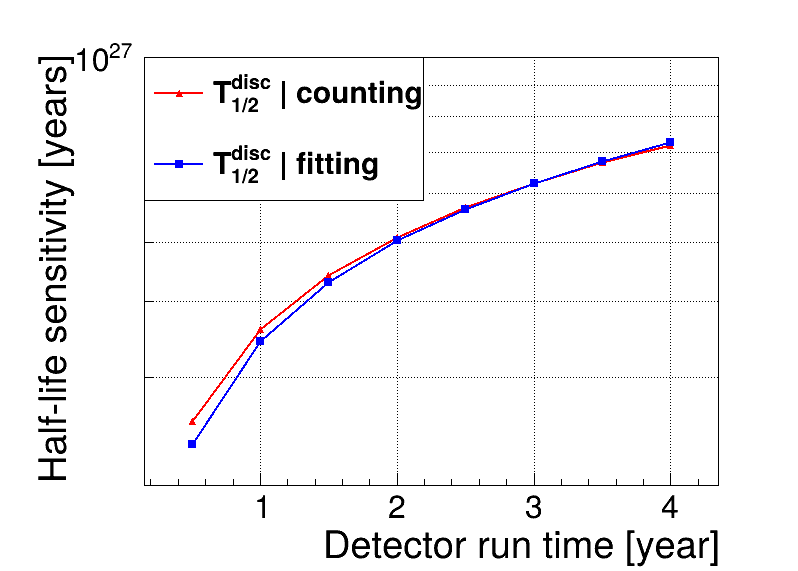}
  \caption{Evolution of the $3\sigma$ discovery sensitivity of the $0\nu\beta\beta$ half-life in $^{130}$Te as a function of detector live time for an energy resolution of 3\%. Red markers show the counting method (using an RoI optimized at this resolution) and blue markers show the spectrum-fitting method. This plot complements Fig.~\ref{fig:crossover} by illustrating the ordering reversal with increasing exposure.}
  \label{fig:reso003_exp_scan}
\end{figure}

\begin{figure}[]
\centering
\includegraphics[width=0.7\linewidth]{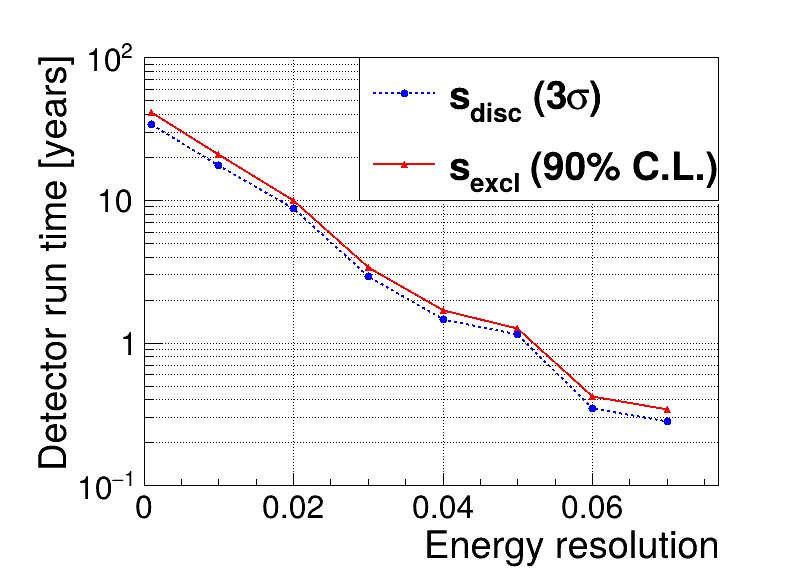}
\caption{Detector run time at which the counting method and the fitting method yield same $3\sigma$ discovery sensitivities (blue line) and 90$\%$ C.L. exclusion sensitivities (red line). The energy resolution ranges from 0.1$\%$ to 7$\%$.}\label{fig:crossover}
\end{figure}

To make this trend more concrete, for a representative resolution of 3$\%$ we explicitly show in Fig.~\ref{fig:reso003_exp_scan} how the $3\sigma$ discovery sensitivity evolves with exposure for both methods. As a result, Figure~\ref{fig:crossover} shows the detector run time at which the counting method and the fitting method yield the same sensitivity.

\begin{figure}[]
\centering
\includegraphics[width=0.7\linewidth]{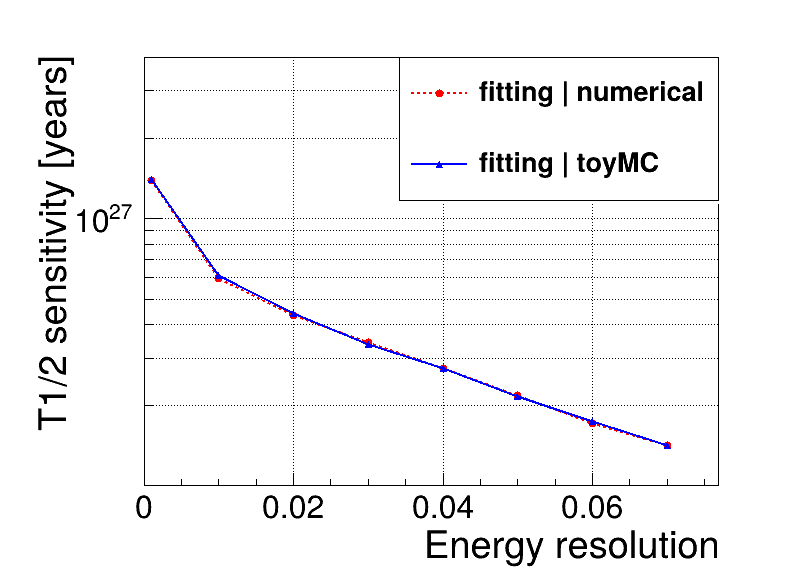}
\caption{The 90$\%$ C.L. exclusion sensitivities of $^{130}$Te $0\nu\beta\beta$ half-life obtained by toy MC simulations described in Section~\ref{subsec:SimulationResult} and the numerical calculation in the same manner as in Section~\ref{subsec:Counting}. The energy resolution ranges from 0.1$\%$ to 7$\%$.
}\label{fig:confirm_theory}
\end{figure}

We also compared the above fitting method sensitivities with the sensitivities numerically calculated in the same manner as in Section~\ref{subsec:Counting}. The resulting sensitivities are shown in Fig.~\ref{fig:confirm_theory}, and the deviation between the numerical calculation and the toy MC sensitivities is less than 1$\%$. This level of deviation is within the acceptable range for our study, indicating the accuracy of our fitting method.

In addition, we conducted thorough simulations under conditions closer to reality. For instance, we modeled natural radioactive backgrounds according to a more specific detector geometry and incorporated them into the background model. The resulting sensitivities are in line with the conclusions of this study, reinforcing the validity of our results.

\section{Discussion: issues in the fitting method}
\label{sec:Discussion}

\subsection{Exposure dependence}
\label{subsec:Fitting_exposure}

As discussed in Ref.~\cite{Cowan:2010js}, in addition to estimating $\sigma_S^2$ using the Asimov dataset value as in Section~\ref{subsec:Counting}, one may also approximate it through the inverse of the Fisher information matrix $I$:  
\begin{equation}
    \sigma^2_S \approx (I^{-1})_{ss}.
\end{equation}

\begin{figure}[]
\centering
\subfigure[]{
\includegraphics[width=0.45\linewidth]{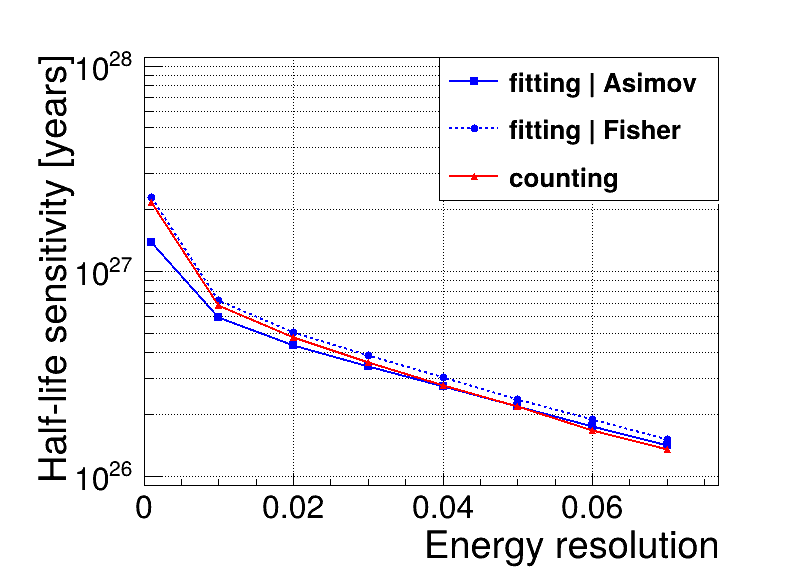}
}
\subfigure[]{
\includegraphics[width=0.45\linewidth]{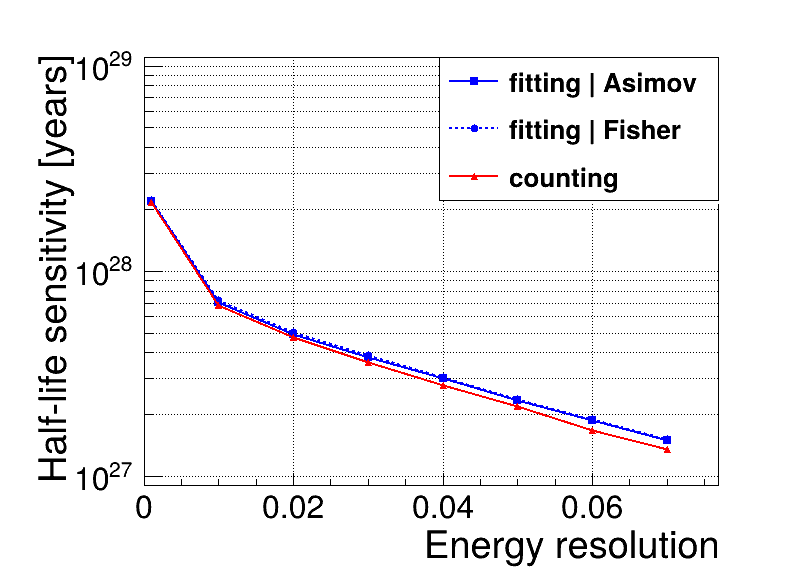}
}
\caption{The 3$\sigma$ discovery sensitivities obtained with a detector run time of (a) one year and (b) 100 years. The blue solid line is the fitting method numerical calculation results using the Asimov dataset approximation, the blue dash line is the fitting method numerical calculation results using the Fisher information matrix approximation, and the red solid line is the counting method results. The energy resolution ranges from 0.1$\%$ to 7$\%$.}\label{fig:compare_exposure}
\end{figure}

The Fisher information matrix establishes the theoretical lower bound on the variance of $S$, and this approximation becomes valid in the large-sample limit, i.e., at high exposure. For instance, Fig.~\ref{fig:compare_exposure} illustrates, under the background conditions defined in Section~\ref{sec:Simulation}, the discovery sensitivity for detector live times of 1 year and 100 years, as estimated with the Fisher information matrix approximation, the Asimov dataset approximation, and the counting method results.

It can be seen that the lower bound on the variance of $S$ inferred from the Fisher information matrix always yields a sensitivity superior to that of the counting method. However, at low exposure this lower bound cannot be saturated, and the actual performance of fitting methods deteriorates. By contrast, at sufficiently high exposure, the results obtained using Asimov dataset values asymptotically converge to those estimated with the Fisher information matrix.

\subsection{Constraint from outside the RoI}
\label{subsec:FittingRoI}

We are particularly interested in how backgrounds outside the signal region contribute to its sensitivity. As a simple example, we consider only one background parameter b, and denote the probability density functions of the signal and background as $f_{\mathrm{sig}}(x)$ and $f_{\mathrm{bkg}}(x)$, respectively.

Using the Fisher information matrix approximation for convenience in the calculation, and in the absence of any penalty terms in $\Delta \chi^2$, one can directly obtain:
\begin{equation}
    \Delta \chi^2_{\alpha}(S=0\mid S=0) = z_{\alpha}^2,
\end{equation}
and
\begin{equation}
    \Delta \chi^2_{\mathrm{median}}(S=0\mid s_0) =  s_0^2 \cdot I_{ss}\mid _{S=0} .
\end{equation}

This is because for the likelihood function $\mathcal{L}(x;s,b) = \frac{sf_{\mathrm{sig}(x)}+bf_{\mathrm{bkg}(x)}}{s+b}$, the elements of the Fisher information matrix satisfy $I_{sb}\mid _{S=0}=I_{bs}\mid _{S=0}=I_{bb}\mid _{S=0}=0$.

Substituting $\mathcal{L}(x; s, b)$ into the definition of Fisher information, we obtain:
\begin{align}
    I_{ss}\mid _{S=0} 
        &= \int b\left( \frac{f_{\mathrm{sig}}(x)}{bf_{\mathrm{bkg}}(x)}-\frac{1}{b} \right)^2 f_{\mathrm{bkg}}(x) dx \notag \\
        &= \int \frac{f_{\mathrm{sig}}(x)^2}{bf_{\mathrm{bkg}}(x)}dx -\frac{1}{b}.
\end{align}

From the above expression, it is clear that background components outside the support of $f_{\mathrm{sig}}(x)$ still contribute to the term $-\frac{1}{b}$. As the amount of background outside the signal region increases, the first term $\int \frac{f_{\mathrm{sig}}(x)^2}{bf_{\mathrm{bkg}}(x)}dx$ would not change since the product $bf_{\mathrm{bkg}}(x)$ within the signal region is held fixed. Consequently, as $b$ becomes sufficiently large, the subtraction term $-1/b$ vanishes, and the Fisher information approaches the value of the first term $\int \frac{f_{\mathrm{sig}}(x)^2}{bf_{\mathrm{bkg}}(x)}dx$.

\subsection{Impact of background shape effect}
\label{subsec:background}

A dedicated study was carried out to assess how background uncertainties affect the sensitivity. The analysis shows that the impact of such uncertainties is generally small and does not alter our main conclusions. Even when unrealistically large uncertainties are assigned to all background components, the resulting degradation in sensitivity remains minor within the large-sample limit.

To further understand the origin of this behavior, we examined how the effect depends on the amount of background overlapping the signal region. For this purpose, we introduced an additional radioactive background component with a strong spectral contribution near the $^{130}$Te signal energy, repeated the sensitivity evaluation, and compared the results with those obtained using a high-$Q$-value isotope. The results confirm that the influence of background uncertainty scales with the background content within the signal region. Only when a dominant background component with both large rate and large uncertainty lies within the signal window does the fitting method become less sensitive than the counting method. In practice, accurate knowledge of the background composition—depending on the specific detector technology—can further mitigate its impact on the sensitivity of the fitting method.

\section{Conclusion}
\label{sec:Conclusion}

In this research, we present a comparison between the counting and fitting methods used in $0\nu\beta\beta$ experiments to evaluate the sensitivity to half-life, considering both discovery sensitivity and exclusion sensitivity. Through statistical derivations, we clarify the underlying statistical principles of both methods and establish their connection via likelihood-based formalism.


A comparative study of two analysis methods—the fitting method and the counting method—was conducted using a simplified detector simulation. The simulation model included background contributions from both two-neutrino double-beta decay and solar neutrino elastic scattering. The results revealed that the fitting method demonstrates superior sensitivity under worse energy resolution conditions, whereas the counting method performs better at better energy resolution. However, over extended operational periods, the fitting method is expected to outperform the counting method eventually. An exposure threshold was identified at which the performance of the two methods becomes equivalent, with this threshold increasing as a function of detector resolution. Validation using Asimov datasets confirmed that the deviation between statistical predictions and simulation outcomes remains below 1$\%$.

The study further demonstrates that, at sufficiently large exposures, the fitting method converges toward the Fisher information matrix approximation. We also show that background events outside the signal region provide valuable spectral information that enhances the performance of the fitting method. In addition, we examined the impact of background uncertainties on sensitivity and found that only when a dominant background component with large uncertainty is present near the signal region, the fitting method can become inferior to the counting method under the large-sample limit.

Overall, this study confirms that the choice between the counting and fitting methods depends on both the exposure and the energy resolution, and should therefore be made with respect to the specific detector configuration. Since this work is based on statistical derivations and toy Monte Carlo simulations, the conclusions are not restricted to the liquid scintillator detector or the energy range considered here. The same methodology can be applied to other types of detectors, such as crystal detectors or time projection chambers, to evaluate sensitivities and guide the choice of method.

\acknowledgments

This work was supported by the National Natural Science Foundation of China (Grant Nos. 12127808 and 12441513).





\bibliographystyle{JHEP}
\bibliography{biblio.bib}

\providecommand{\href}[2]{#2}\begingroup\raggedright\begin{thebibliography}{10}

\bibitem{Dolinski:2019nrj}
M.J.~Dolinski, A.W.P.~Poon and W.~Rodejohann, \emph{Neutrinoless double-beta
  decay: Status and prospects}, {\emph{Ann. Rev. Nucl. Part. Sci.} {\bfseries
  69} (2019) 219} [\href{https://arxiv.org/abs/1902.04097}{{\ttfamily
  1902.04097}}].

\bibitem{Rodejohann:2011mu}
W.~Rodejohann, \emph{Neutrino-less double beta decay and particle physics},
  {\emph{Int. J. Mod. Phys. E} {\bfseries 20} (2011) 1833}
  [\href{https://arxiv.org/abs/1106.1334}{{\ttfamily 1106.1334}}].

\bibitem{Zeng:2016pnx}
Z.~Zeng et~al., \emph{Characterization of a broad-energy germanium detector for
  its use in cjpl}, {\emph{Nucl. Sci. Tech.} {\bfseries 28} (2017) 7}.

\bibitem{CUORE:2022fgm}
{\scshape CUORE} collaboration, \emph{Latest results from the cuore
  experiment}, {\emph{J. Low Temp. Phys.} {\bfseries 209} (2022) 927}.

\bibitem{CUPID:2020aow}
{\scshape CUPID} collaboration, \emph{New limit for neutrinoless double-beta
  decay of $^{100},$mo from the cupid-mo experiment}, {\emph{Phys. Rev. Lett.}
  {\bfseries 126} (2021) 181802}
  [\href{https://arxiv.org/abs/2011.13243}{{\ttfamily 2011.13243}}].

\bibitem{Chen:2022rzg}
W.~Chen, L.~Ma, J.-H.~Chen and H.-Z.~Huang, \emph{Gamma-, neutron-, and
  muon-induced environmental background simulations for $^{100},$mo-based
  bolometric double-beta decay experiment at jinping underground laboratory},
  {\emph{Nucl. Sci. Tech.} {\bfseries 34} (2023) 135}
  [\href{https://arxiv.org/abs/2206.09288}{{\ttfamily 2206.09288}}].

\bibitem{NnDEx-100:2023alw}
{\scshape N\ensuremath{\nu},DEx-100} collaboration,
  \emph{N\ensuremath{\nu},dex-100 conceptual design report}, {\emph{Nucl. Sci.
  Tech.} {\bfseries 35} (2024) 3}
  [\href{https://arxiv.org/abs/2304.08362}{{\ttfamily 2304.08362}}].

\bibitem{GERDA:2020xhi}
{\scshape GERDA} collaboration, \emph{Final results of gerda on the search for
  neutrinoless double-$\beta$ decay}, {\emph{Phys. Rev. Lett.} {\bfseries 125}
  (2020) 252502} [\href{https://arxiv.org/abs/2009.06079}{{\ttfamily
  2009.06079}}].

\bibitem{CDEX:2022bdk}
{\scshape CDEX} collaboration, \emph{Search for neutrinoless double-beta decay
  of ge76 with a natural broad energy germanium detector}, {\emph{Phys. Rev. D}
  {\bfseries 106} (2022) 032012}
  [\href{https://arxiv.org/abs/2205.10718}{{\ttfamily 2205.10718}}].

\bibitem{PandaX:2023ggs}
{\scshape PandaX} collaboration, \emph{Searching for two-neutrino and
  neutrinoless double beta decay of xe134 with the pandax-4t experiment},
  {\emph{Phys. Rev. Lett.} {\bfseries 132} (2024) 152502}
  [\href{https://arxiv.org/abs/2312.15632}{{\ttfamily 2312.15632}}].

\bibitem{KamLAND-Zen:2022tow}
{\scshape KamLAND-Zen} collaboration, \emph{Search for the majorana nature of
  neutrinos in the inverted mass ordering region with kamland-zen},
  {\emph{Phys. Rev. Lett.} {\bfseries 130} (2023) 051801}
  [\href{https://arxiv.org/abs/2203.02139}{{\ttfamily 2203.02139}}].

\bibitem{Ji:2019yca}
X.~Ji et~al., \emph{Combined neyman\textendash{},pearson chi-square: An
  improved approximation to the poisson-likelihood chi-square}, {\emph{Nucl.
  Instrum. Meth. A} {\bfseries 961} (2020) 163677}
  [\href{https://arxiv.org/abs/1903.07185}{{\ttfamily 1903.07185}}].

\bibitem{Cowan:2010js}
G.~Cowan et~al., \emph{{Asymptotic formulae for likelihood-based tests of new
  physics}}, {\emph{Eur. Phys. J. C} {\bfseries 71} (2011) 1554}
  [\href{https://arxiv.org/abs/1007.1727}{{\ttfamily 1007.1727}}].

\bibitem{DellOro:2016tmg}
S.~Dell'Oro et~al., \emph{{Neutrinoless double beta decay: 2015 review}},
  {\emph{Adv. High Energy Phys.} {\bfseries 2016} (2016) 2162659}
  [\href{https://arxiv.org/abs/1601.07512}{{\ttfamily 1601.07512}}].

\bibitem{Zuber:2020kci}
K.~Zuber, \emph{Neutrino Physics}, Taylor \& Francis, Boca Raton (2020).

\bibitem{Simkovic:2018rdz}
F.~{\v{S}}imkovic, R.~Dvornick{\'y}, D.~Stef{\'a}nik and A.~Faessler,
  \emph{{Improved description of the $2\nu\beta\beta$ -decay and a possibility
  to determine the effective axial-vector coupling constant}},
  \href{https://doi.org/10.1103/PhysRevC.97.034315}{\emph{Phys. Rev. C}
  {\bfseries 97} (2018) 034315}
  [\href{https://arxiv.org/abs/1804.04227}{{\ttfamily 1804.04227}}].

\bibitem{Serenelli:2016dgz}
A.~Serenelli, \emph{Alive and well: a short review about standard solar
  models}, {\emph{Eur. Phys. J. A} {\bfseries 52} (2016) 78}
  [\href{https://arxiv.org/abs/1601.07179}{{\ttfamily 1601.07179}}].

\bibitem{Bilenky:2005mx}
S.M.~Bilenky, \emph{Neutrino masses, mixing and oscillations}, {\emph{Prog.
  Part. Nucl. Phys.} {\bfseries 57} (2006) 61}
  [\href{https://arxiv.org/abs/hep-ph/0510175}{{\ttfamily hep-ph/0510175}}].

\bibitem{Baudis:2013qla}
L.~Baudis et~al., \emph{Neutrino physics with multi-ton scale liquid xenon
  detectors}, {\emph{JCAP} {\bfseries 01} (2014) 044}
  [\href{https://arxiv.org/abs/1309.7024}{{\ttfamily 1309.7024}}].

\end{thebibliography}\endgroup

\end{document}